\documentclass[preprints,article,accept,moreauthors,pdftex]{Definitions/mdpi}

\firstpage{1} 
\pubvolume{1}
\issuenum{1}
\articlenumber{0}
\pubyear{2023}
\copyrightyear{2023}
\datereceived{} 
\dateaccepted{} 
\datepublished{} 
\hreflink{https://doi.org/} 
\pdfoutput=1

\Title{AN EXACT MODEL OF A GRAVITATIONAL WAVE IN THE BIANCHI III UNIVERSE BASED ON SHAPOVALOV II WAVE SPACETIME AND THE QUADRATIC THEORY OF GRAVITY}

\TitleCitation{An exact model of a gravitational wave in the Bianchi III universe based on Shapovalov II wave spacetime and the quadratic theory of gravity
}

\Author{
Konstantin Osetrin $^{1, 2,
\ddagger}$\orcidA{0000-0003-4739-2788}, 
Ilya Kirnos $^{1,\ddagger}$ \orcidB{0000-0003-3737-3583}, 
and 
Evgeny Osetrin $^{1,3,\ddagger}$ \orcidC{0000-0001-5998-2238}
}

\AuthorNames{Konstantin Osetrin, Ilya Kirnos and Evgeny Osetrin}

\AuthorCitation{Osetrin, K.E.; Kirnos, I.V.; Osetrin, E.K.}

\address{%
$^{1}$ \quad Tomsk State Pedagogical University; osetrin@tspu.edu.ru\\
$^{2}$ \quad Tomsk State University; osetrin@gmail.com\\
$^{3}$ \quad Tomsk State University of Control Systems and Radioelectronics; osetrin@mail.ru\\
}

\secondnote{These authors contributed equally to this work.}

\abstract{
Exact models of primordial gravitational waves in the Bianchi type III universe are constructed on the basis of the quadratic theory of gravity with a scalar field and pure radiation in Shapovalov wave spacetimes of type II (subtype~2). Exact solutions of field equations and scalar equation are obtained. The characteristics of pure radiation are determined. An explicit form of the scalar field functions included in the Lagrangian of the considered quadratic theory of gravity is found.
Trajectories of propagation of light rays in the considered gravitational-wave models are obtained.
}

\keyword{gravitational waves; Bianchi universes; Shapovalov wave spacetimes; quadratic theories of gravity; scalar field; pure radiation} 

\MSC{83C10, 83C35}
\begin{document}
%

\section{Introduction}

At present, after receiving an array of observational data from satellite telescopes on the CMB (electromagnetic  cosmic microwave background of the universe) and revealing the nonisotropy of this background \cite{Bennett2013}, the urgency of the problem of studying the influence on the primary plasma and on the microwave background from primordial gravitational waves has increased. Primordial gravitational waves could also influence the formation of early inhomogeneities and primordial black holes \cite{Saito200916,Saito2010867}. Primordial gravitational waves arose in a situation where quantum corrections of the gravitational equations played a significant role, which can also be reflected in the characteristics of the microwave background that we observe. These corrections, in the second approximation, can be expressed as terms quadratic in curvature in the Lagrangian of the gravitational field \cite{Boulware-Deser}. The inflationary model of the early universe is also based on the existence of the inflaton scalar field, which plays a significant role in the initial stages of the evolution of the universe \cite{Guth}. Therefore, the study of homogeneous non-isotropic models of gravitational waves in theories of gravity quadratic in scalar curvature with a scalar field and further evaluation of their influence on the primordial plasma, the microwave background of the universe, and the formation of primary inhomogeneities is an urgent, significant, but difficult task.

In papers \cite{Osetrin2022894,Osetrin2022EPJP856}, exact models of gravitational waves in the Bianchi universes were obtained on the basis of Shapovalov wave spaces, for which particle trajectories and exact solutions of geodesic deviation equations were found.

Modified gravity theories are used in modeling various astrophysical phenomena. Paper \cite{Bamba-Capozziello-Nojiri-Odintsov} reviews the modifications of general relativity and compares them with cosmographic tests. In addition, \cite{Nojiri-Odintsov, Nojiri-Odintsov-Oikonomou} gives a large overview of theories with Lagrangians of the form $F(R)$, theories containing the Gauss-Bonnet term $F(R,\mathcal{G}) $ and scalar-tensor theories, including the description with the help of such theories of the stage of inflation and the current stage of the accelerated expansion of the universe.

As for gravitational waves emitted by close binary systems, their features arising in various modifications of general relativity and comparison of these features with observational data obtained by the LIGO/Virgo/KAGRA collaboration are discussed in \cite{Zhu-Zhao-Yan-Gong-Wang}. Gravitational waves arising from the merger of binary neutron stars are considered in accordance with the Einstein-Gauss-Bonnet theory with a scalar field in \cite{East-Pretorius}.

In our work, we develop an approach to the study of gravitational waves based 
on Shapovalov wave spacetimes \cite{Osetrin2020Symmetry,Osetrin2021Sym1173,Osetrin2022Sym14122636,Osetrin2022Universe8120664}
and consider the generalized quadratic theory of gravity with a scalar field, for which the authors have previously obtained an explicit form of the field equations \cite{Osetrin2022Universe8120664}. A model of a gravitational wave in the Bianchi type III universe, which belongs to the class of Shapovalov type II wave spacetimes, is considered.

Shapovalov spacetimes in privileged coordinate systems allow separation of variables in the eikonal equation and in the Hamilton-Jacobi equation for test particles with separation of wave variables, along which the spacetime interval vanishes, which plays an important role in obtaining exact solutions in these models. As is known from recent observations on the detection of gravitational waves and electromagnetic radiation from neutron star merger \cite{Abbott2017PRL161101}, the gravitational-wave and electromagnetic signals reached the observers almost simultaneously, which made it possible to experimentally confirm the equality of the propagation velocity of gravitational waves and the speed of light. Therefore, Shapovalov spacetimes with separable wave variables have a physical interpretation as wave models of spacetime and serve to describe the propagation of gravitational-wave disturbances. Type II Shapovalov spacetimes allow two commuting Killing vectors and include two non-ignorable variables on which the metric in privileged coordinate systems depends, which for gravity theories with a scalar field provides additional non-trivial possibilities for studying the behavior of a scalar field and obtaining admissible analytical forms for scalar field functions, included in the Lagrangian of the theory, including the potential of the scalar field, which models the behavior of the ''cosmological constant'' (modeling the phenomenon of ''dark energy'').

We also note that Shapovalov spacetimes allow one to obtain exact solutions of the scalar equation, which is part of the field equations of scalar-tensor gravity theories, by the method of separation of variables in privileged coordinate systems.

\section{Shapovalov wave spacetimes and the Bianchi type III universe}

In this paper, as the spacetime model under study, we will use the Shapovalov type II wave spacetime model (B2 subtype) for Bianchi type III universes. The choice of this model is due to the fact that this model refers to gravitational-wave spacetimes, the metric of which in the privileged coordinate system is dependent on two variables and has Bianchi type III symmetries of the universe. Thus, the problem under consideration, with its relative simplicity, has the necessary properties for constructing and studying an accurate model of the primordial gravitational wave, i.e. makes it possible to construct exact solutions of the field equations of the quadratic theory of gravity and to find the analytical form of the scalar field functions entering into the Lagrangian of the theory.

In a privileged coordinate system, where the eikonal equation and the Hamilton-Jacobi equation for test particles allow integration by the method of separation of variables, the metric of the considered spacetime can be written in the following form:
\begin{equation} 
{\rm g}_{\mu\nu}(x^0,x^3)= \left(                                                            
              \begin{array}{cccc}
               0 & \frac{1}{{x^3}^2} & 0 & 0 \\
               \frac{1}{{x^3}^2} & 0 & 0 & 0 \\
               0 & 0 & \frac{{x^0}^{{\alpha}}}{{x^3}^2} & 0 \\
               0 & 0 & 0 & \frac{1}{{x^3}^2} \\
              \end{array}
              \right)
,\qquad
\mu, \nu=0,1,2,3
\label{MetricD}
,\end{equation}
where ${\alpha}$ is the constant parameter of the model, the variable $x^0$ is the wave variable along which the spacetime interval vanishes.

The metric determinant has the following form:
\begin{equation} 
{g=\det g_{\mu\nu}=}-\frac{\left({x^0}\right)^{{\alpha}}}{\left({x^3}\right)^8}
.\end{equation}
The origin is a singular point of the preferred coordinate system at which the coordinate system gives a singularity.

The spacetime model under consideration admits a Bianchi type III homogeneity symmetry subgroup and has type N according to Petrov's classification.
Weyl conformal curvature tensor $C_{\mu\nu\alpha\beta}$ has two nonzero components
\begin{equation} 
{C}_{0202} = -\frac{{\alpha} ({\alpha}-2) {x^0}^{{\alpha}-2}}{8 {x^3}^2}
,\end{equation}
\begin{equation} 
{C}_{0303} = \frac{{\alpha} ({\alpha}-2)}{8 {x^0}^2 {x^3}^2}
.\end{equation}
where ${\alpha}$ is the constant parameter of the model, the variable $x^0$ is the wave variable along which the spacetime interval vanishes.

The Weyl tensor vanishes identically and the spacetime becomes conformally flat at $\alpha=0$ and at $\alpha=2$. The Riemann curvature tensor $R_{\mu\nu\alpha\beta}$ identically does not vanish for any model parameters.

\section{Quadratic theory of gravity with scalar field}

We will consider quadratic theory of gravity with scalar field with the following Lagrangian
\begin{equation} 
 L=\frac{1}{2\kappa^2}\Bigl(\sigma({\Phi})R+\gamma({\Phi})R^2\Bigr)
- \xi({\Phi})\,G
-\frac{\omega}{2} g^{\mu\nu}\partial_\mu{\Phi}\partial_\nu{\Phi}- V({\Phi})
+{L}_{Matter}
\label{Lagrangian}
.\end{equation}
Here $\kappa$ is Newton's constant, $R$ is the scalar curvature, $G$ is the Gauss-Bonnet term, ${\Phi}$ is the scalar field, $V({\Phi})$ is the scalar field potential modeling including the phenomenon of "dark energy" and the cosmological constant $\Lambda$, the parameter ${\omega}$ is a constant, the functions $\sigma({\Phi})$, $\gamma({\Phi})$, $\xi({\Phi})$ and $V({\Phi})$ are assumed to be arbitrary functions of the scalar field, whose analytical form must be obtained by solving the field equations of the theory.

As the energy-momentum tensor, we will use the pure radiation tensor, which can model both high-energy electromagnetic radiation and the high-frequency part of gravitational radiation:
\begin{equation} 
T_{\alpha\beta}=\varepsilon L_{\alpha}L_{\beta},
\qquad
g^{\alpha\beta}L_{\alpha}L_{\beta}=0
\label{TEIR}
,\end{equation}
where $\varepsilon(x^\alpha)$ is the energy density of pure radiation, $L_{\alpha}$ is the wave vector of pure radiation.

Note for reference that, for the spacetimes under consideration, Einstein's equations of general relativity with the energy-momentum tensor of pure radiation (\ref{TEIR}), due to the positiveness of the radiation energy $\varepsilon$, lead to a condition on the constant parameter of the model $\alpha$ of the form $0\le \alpha\le 2$.

By direct variation of the Lagrangian (\ref{Lagrangian}) with respect to the metric $g_{\mu\nu}$ and the scalar field ${\Phi}$, in \cite{Osetrin2022Universe8120664} the authors obtained the following form of the field equations of the theory under consideration:
\begin{equation}
Q_{\mu\nu}=T_{\mu\nu}
,\qquad
\alpha,\beta,\gamma,\mu,\nu=0,...3
\label{FieldEquations1}
,\end{equation}
\begin{eqnarray}
Q_{\mu\nu}&\equiv&
\displaystyle\frac{1}{\kappa^2}
\Bigl(
\left[\sigma(\Phi)+2\gamma(\Phi)R\right]R_{\mu\nu}
-\nabla_\mu\nabla_\nu\left[\sigma(\Phi)
+2\gamma(\Phi)R\right]
\nonumber\\&&
\vphantom{\frac{1}{2}}
+g_{\mu\nu}\Box\left[\sigma(\Phi)
+2\gamma(\Phi)R\right]
\Bigr)
-8\nabla_\alpha\nabla_\beta\left[\xi(\Phi) R_{\mu\phantom{\alpha}\nu}^{\phantom{\mu}\alpha\phantom{\nu}\beta}\right]
\nonumber\\&&
\vphantom{\frac{1}{2}}
- 8\nabla_\gamma\nabla_\nu\left[\xi(\Phi){R^\gamma}_\mu\right]
-8\nabla_\gamma\nabla_\mu\left[\xi(\Phi){R^\gamma}_\nu\right]
\nonumber\\&&
\vphantom{\frac{1}{2}} 
+ 8\Box\left[\xi(\Phi)R_{\mu\nu}\right]
+ 8g_{\mu\nu}\nabla_\alpha\nabla_\beta\left[\xi(\Phi)R^{\alpha\beta}\right]
\nonumber\\&&
\vphantom{\frac{1}{2}}
+ 4\nabla_\mu\nabla_\nu\left[\xi(\Phi)R\right]
-4g_{\mu\nu}\Box\left[\xi(\Phi)R\right]
\nonumber\\&&
\vphantom{\frac{1}{2}}
+16\xi(\Phi)R_{\alpha\mu}{R^\alpha}_\nu
- 4\xi(\Phi)RR_{\mu\nu}
- 4\xi(\Phi)R_{\mu\alpha\beta\gamma}{R_\nu}^{\alpha\beta\gamma}
\nonumber\\&&
\vphantom{\frac{1}{2}}
-\omega\partial_\mu{\Phi}\partial_\nu{\Phi}
+\frac{\omega}{2} g_{\mu\nu} g^{\alpha\beta}\partial_\alpha{\Phi}\partial_\beta{\Phi}
\nonumber\\&&
- \displaystyle\frac{\sigma(\Phi)R
+\gamma(\Phi)R^2}{2\kappa^2}g_{\mu\nu}
+ \xi(\Phi)\mathcal{G}g_{\mu\nu}
+V(\Phi)g_{\mu\nu} 
\label{FieldEquations2}
,
\end{eqnarray}
where $\nabla_\mu$ is the covariant derivative,
$\Box=g^{\alpha\beta}\nabla_\alpha\nabla_\beta$ is the d'Alembert operator, $T_{\mu\nu}$ is the energy-momentum tensor.

The scalar equation takes the following form:
\begin{equation}
\omega\,\Box{\Phi}+\displaystyle\frac{\sigma'(\Phi)R+\gamma'(\Phi)R^2}{2\kappa^2}
- \xi'(\Phi)\,\mathcal{G}-V'(\Phi)=0
\label{ScalarEquation}
,\end{equation}
where we denote the ordinary derivative of one variable with a prime.

Investigation of the compatibility of field equations (\ref{FieldEquations1}) with the energy-momentum tensor of pure radiation (\ref{TEIR}) for the metric (\ref{MetricD}) in the general case when the scalar field depends on all four variables $\Phi= \Phi(x^0,x^1,x^2,x^3)$ leads to inconsistency of the field equations. It follows from the field equations that in the considered problem in the privileged coordinate system the scalar field cannot depend on the ignored variables $x^1$ and $x^2$, on which the spacetime metric does not depend. Therefore, below we will consider cases where the scalar field depends only on non-ignorable variables: the wave variable $x^0$ and the non-isotropic (non-wave) variable $x^3$.

\section{Case I. The scalar field depends on the wave variable $x^0$ only}

Let us consider the case when the scalar field depends on only one wave variable $x^0$, i.e. $\Phi={\phi_0}(x^0)$. In this case, the d'Alembert operator of the scalar field vanishes, and from the scalar equation we obtain the condition:
\begin{equation} 
\frac{d}{d\Phi}
\left(
-\frac{72 }{\kappa ^2}{\gamma}\left( \Phi \right)
+\frac{6 }{\kappa^2}{\sigma}\left( \Phi \right)
+84 {\xi}\left( \Phi \right)
+{V}\left( \Phi \right)
\right)
=0
.\end{equation}
The field equations give the following relations for the components of the wave vector $L_\mu$ of pure radiation:
\begin{equation}
L_0\ne0
,\qquad
L_1=L_2=L_3=0
.\end{equation}
Thus, pure radiation for the model under consideration propagates in a privileged coordinate system along the coordinate $x^1$, i.e. $L^\beta=\bigl\{0,L^1,0,0\bigr\}$. The remaining field equations give the following relations:
$$
\kappa ^2 
{\varepsilon}{L_0}^2
=
\frac{{\alpha}({\alpha}-2)  (24 {\gamma}({\Phi}{})-{\sigma}({\Phi}{}))}{4  {x^0}^2}
$$
\begin{equation} 
\mbox{}
-
{\phi_0}'{}^2 
\left(
\kappa ^2 \left(8 {\xi}''{}+{\omega}\right)-24 {\gamma}''{}+{\sigma}''{}
\right)
-
{\phi_0}''{} \left(-24 {\gamma}'{}+{\sigma}'{}+8 \kappa ^2 {\xi}'{}\right)
,\end{equation}
\begin{equation} 
({3 }/{\kappa ^2}){\sigma}({\Phi}{})+60 {\xi}({\Phi}{})
+{V}({\Phi}{})
=0
,\end{equation}
\begin{equation} 
{\phi_0}'{} 
\frac{d}{d\Phi}
\Bigl(
{\sigma}\left( \Phi \right)
+8 \kappa ^2 {\xi}\left( \Phi \right)
-24 {\gamma}\left( \Phi \right)
\Bigr)
=0
.\end{equation}

Expressing the function $\xi(\Phi )$ and the scalar potential $V(\Phi )$ from these remaining equations, we obtain
\begin{equation} 
\xi(\Phi ) = \frac{3 {\gamma}({\Phi}{})}{\kappa ^2}-\frac{{\sigma}({\Phi}{})}{8 \kappa ^2}+\frac{{\Lambda}}{24}
,\qquad
{\Lambda} - const
\label{X0xi}
,\end{equation}
\begin{equation} 
V(\Phi ) = -\frac{360 {\gamma}({\Phi}{})-9 {\sigma}({\Phi}{})+5 {\Lambda} \kappa ^2}{2 \kappa ^2}
\label{X0V}
,\end{equation}
where ${\Lambda}$ is the integration constant related to the cosmological constant. At the same time, the functions
${\sigma}({\Phi})$ and ${\gamma}({\Phi})$ remain arbitrary.

Taking into account the relations (\ref{X0xi})-(\ref{X0V}), from the entire set of field equations, one relation remains, including the term for pure radiation with energy density ${\varepsilon}(x^0)$:
\begin{equation} 
{\varepsilon}{}{L_0}^2
=
\frac{
{\alpha}({\alpha}-2)
}{4 \kappa ^2 {x^0}^2}
\,
\Bigl(
24 {\gamma}({\phi}_0)-{\sigma}({\phi}_0)
\Bigr)
-
{\omega}{{\phi_0}'}^2 
\label{PureRadiationX0}
.\end{equation}

If we assume that pure radiation is external and is given, then
this equation is used to define the scalar field ${\Phi}={\phi}_0(x^0)$. Those. in this version, external pure radiation (electromagnetic, gravitational, etc.) ''generates'' a scalar field. If we consider the interpretation of pure radiation as a high-frequency component of the gravitational wave itself on the basis of the considered metric, when there is no external radiation, then the relation (\ref{PureRadiationX0}) gives us the energy density ${\varepsilon}(x^0)$ of the high-energy component of the gravitational wave in the high frequency range. In both versions, pure radiation propagates in a privileged coordinate system along the variable $x^1$, i.e. $L^{\beta}=\{0,L^1,0,0\}$.

If we fix the choice of the scalar potential $V(\Phi )$ as a cosmological constant, then one arbitrary function of the scalar field remains, which can also be fixed by imposing an additional condition, such as the constant coefficient $\sigma$ under scalar curvature in the Lagrangian of the theory ${\sigma}({\Phi}{})=\mbox{const}$.
In this case, all coefficients in the action of the considered model are determined and converted into constants:
\begin{equation} 
\gamma 
= 
\frac{1}{360} \left(-5 \kappa ^2 \Lambda -2 \kappa ^2 V+9 \sigma \right)
,\end{equation}
\begin{equation} 
\xi 
=
\frac{5 \kappa ^2 \Lambda +\kappa ^2 V-12 \sigma}{60 \kappa ^2}
.\end{equation}

Note that the obtained solution in the considered version, when the scalar field depends in the privileged coordinate system only on one wave variable $x^0$, can have a narrowly ''special'' form. Therefore, to clarify the form of the scalar field functions in the Lagrangian of the considered modified theory of gravity, it is important to obtain variants of the exact solution when the scalar field depends on several variables and, in this case, it is possible to obtain an explicit analytical form of the scalar field functions.

We also note that when solving complex field equations of quadratic gravity theories numerically, it is possible to obtain an analytical form of scalar field functions only approximately in the form of a finite series. Obtaining exact solutions with an explicit analytical form of scalar field functions allows us to determine, among other things, which scalar field functions can be included in such expansions in the numerical analysis of more complex gravitational models in quadratic gravity theories.

\section{Case II. The scalar field depends on non-isotropic variable $x^3$ only}

In this section, we consider the solution of field equations for the case when the scalar field depends on only one non-ignorable non-isotropic (non-wave) variable~$x^3$:
\begin{equation}
\Phi= {\phi_3}(x^3)
.\end{equation}

To simplify the field equations, instead of the function $\xi(\Phi )$ and the scalar potential $V(\Phi )$, we introduce new auxiliary functions ${{Z}}(\Phi )$ and ${{Y}}(\Phi )$ according to the rules:
$$
\xi(\Phi ) = \frac{1}{48} \left(
6 \,
\frac{
24 {\gamma}(\Phi)-{\sigma}(\Phi)
}{\kappa ^2}
+{{Y}}(\Phi)\right)
,$$
$$
V(\Phi ) =\frac{9 {\sigma}(\Phi)}{2 \kappa ^2} -\frac{180 {\gamma}(\Phi)}{\kappa ^2}+{{Z}}(\Phi)
-\frac{7}{4} {{Y}}(\Phi)
.$$

Then the scalar equation takes the form:
\begin{equation}
{\omega} {x^3} \left(
{x^3} {\phi_3}''{}-2 {\phi_3}'{}
\right)
=
{{Z}}'(\Phi)
.\end{equation}

The remaining three field equations that do not turn into identities can be represented in the following compact form:
$$
{\alpha} 
({\alpha}-2) 
\biggl[
{x^3} 
\Bigl(
-{x^3} {\phi_3}''{} \left(144 {\gamma}'{}-6 {\sigma}'{}+\kappa ^2 {{Y}}'{}\right)
$$
$$
-{\phi_3}'{} \left({x^3} {\phi_3}'{} \left(144 {\gamma}''{}-6 {\sigma}''{}+\kappa ^2 {{Y}}''{}\right)+144 {\gamma}'{}-6 {\sigma}'{}+\kappa ^2 {{Y}}'{}\right)
\Bigr)
$$
\begin{equation}
+144 {\gamma}{}-6 {\sigma}{}
\biggr]
=
{24 \kappa ^2 {x^0}^2}
{L_0}^2 {\varepsilon}{}
\label{x3E00}
.\end{equation}
\begin{equation}
{x^3} 
\left[
{x^3} {\phi_3}'{}^2 \left(3 {\omega}+{{Y}}''{}\right)+{{Y}}'{} \left({x^3} 
{\phi_3}''{}-{\phi_3}'{}\right)
\right]
=
-3
\Bigl(
2{{Z}}{}- {{Y}}{}
\Bigr)
\label{x3E22}
.\end{equation}
\begin{equation}
{x^3} {\phi_3}'{} \left({\omega} {x^3} {\phi_3}'{}+{{Y}}'{}\right)
=
2 {{Z}}{}-{{Y}}{}
\label{x3E33}
.\end{equation}

Substitute ${{Y}}'$ from the equation (\ref{x3E33}) into the equation (\ref{x3E22}), we get the relation:
\begin{equation}
{x^3} {\phi_3}'{}^2 \left(3 {\omega}+{{Y}}''({\phi_3}{})\right)+{{Y}}'({\phi_3}{}) \left({x^3} 
{\phi_3}''{}-{\phi_3}'{}\right)
=
-3
{\phi_3}'{} \left({\omega} {x^3} {\phi_3}'{}+{{Y}}'({\phi_3}{})\right)
.\end{equation}

The compatibility condition for the equations (\ref{x3E22}) and (\ref{x3E33}) gives the following effect:
$$
{x^3} {\phi_3}'{}^2 \left(6
   {\omega}+{{Y}}''({\phi_3}{})\right)+{{Y}}'({\phi_3}{})
   \left({x^3} {\phi_3}''{}+2 {\phi_3}'{}\right)
   =0.
$$

Thus, we obtain the following relations for the functions ${{Y}}$ and ${{Z}}$:
\begin{equation}
{{Y}}''({\phi_3}{})=
-6 {\omega}
-
{{Y}}'({\phi_3}{}) \,
\frac{\left({x^3} {\phi_3}''{}+2
   {\phi_3}'{}\right)}{{x^3} {\phi_3}'{}^2}
\label{Theta2}
.\end{equation}
\begin{equation}
{{Z}}({\phi_3}{})=
\frac{
{x^3} {\phi_3}'{} 
}{2}
\Bigl(
{\omega} {x^3} {\phi_3}'{}+{{Y}}'({\phi_3}{})
\Bigr)
+
\frac{1}{2}\,
{{Y}}({\phi_3}{})
\label{Psi}
.\end{equation}
\begin{equation}
{{Z}}'({\phi_3}{})
=
{\omega} {x^3} \left({x^3} {\phi_3}''{}-2 {\phi_3}'{}\right)
%
\label{Psi1}
.\end{equation}

Compatibility condition for the equations (\ref{Psi}) and (\ref{Psi1}), taking into account the equation (\ref{Theta2})
gives the corollary that ${{Y}}$ is a constant. But then $\omega=0$ follows from (\ref{Theta2}), which leads to a contradiction.

Thus, in the problem we are considering, the scalar field $\Phi$ cannot depend on only one non-isotropic (non-wave) variable $x^3$, although this variable is not ignored and the spacetime metric depends on this variable.

\section{Case III. 
The scalar field depends on all non-ignored variables
}

We call variables ''ignorable'' if in the privileged coordinate system under consideration the spacetime metric depends on them and ''ignored'' variables if the spacetime metric does not depend on these variables. In this section, we consider the case where the scalar field depends on two non-ignorable variables: the wave variable $x^0$ and the non-isotropic variable $x^3$. In this case, the field equations have a more complex and cumbersome form. We will consider the case when the scalar field can be written in a ''separated'' form:
\begin{equation} 
{\Phi}={\phi_0}(x^0)\,{\phi_3}(x^3)
.\end{equation}

Then the separation of variables in the scalar equation gives the relationship:
\begin{equation} 
\phi_3{}'' = \frac{K {\phi_3}}{{\omega} {x^3}^2}+\frac{2 {\phi_3}'{}}{{x^3}}
,\qquad
K=\mbox{const}
\label{phi3x0x3}
,\end{equation}
\begin{equation} 
{V}(\Phi) = {\Lambda}+\frac{72 {\gamma}({\Phi})-6 {\sigma}({\Phi})}{\kappa ^2}-84 {\xi}({\Phi})+\frac{K {\Phi}^2}{2}
\label{Vx0x3}
,\end{equation}
where $K$ is the separation constant, $\Lambda$ is the integration constant in the scalar equation related to the cosmological constant.

The study of the compatibility of the remaining field equations leads to relations for the components of the wave vector of pure radiation of the form:
\begin{equation}
L_0\ne 0
,\qquad
L_1=0
,\qquad
L_2=0
,\qquad
L_3=0
\label{Lx0x3}
.\end{equation}
Using the relations (\ref{phi3x0x3}), (\ref{Vx0x3}) and (\ref{Lx0x3}) makes it possible to simplify the field equations and write them in the following compact form:

$$
4 {\Phi}{} 
\,
\Biggl(
\frac{2 {\alpha}  ({\alpha}-2) \kappa ^2 \left({\omega} {x^3}^2 {\phi_3}'{}^2 {\Phi}{} {\xi}''{}+{\phi_3} \left(3 {\omega} {x^3} {\phi_3}'{}+K {\phi_3}\right) {\xi}'{}\right)}{{\omega} {{\phi_3}(x^3)}^2}
$$
$$
+\frac{{x^0}^2 {\phi_0}'{}^2 {\Phi}{} \left(\kappa ^2 \left(8 {\xi}''{}+{\omega}\right)-24 {\gamma}''{}+{\sigma}''{}\right)}{{{\phi_0}({x^0})}^2}
+\frac{{x^0}^2 {\phi_0}'' \left(
8 \kappa ^2 {\xi}'{}
-24 {\gamma}'{}+{\sigma}'{}
\right)
}{{\phi_0}({x^0})}
\Biggr)
$$
\begin{equation} 
\mbox{}
+{\alpha} ({\alpha}-2) {\sigma}{}
-
24{\alpha} ({\alpha}-2) {\gamma}{}
+ 
{4 \kappa ^2 {x^0}^2}
{\varepsilon}{L_0}^2
=0
\label{E00x0x3}
,\end{equation}
$$
{\phi_3}^2 
\Bigl(
2 {\omega} \kappa ^2 \left(
{\Lambda}
-24 {\xi}{}
\right)+144 {\omega} {\gamma}{}-6 {\omega} {\sigma}{}+{\omega} K \kappa ^2 {\Phi}{}^2+2 K {\Phi}{} \left(
8 \kappa ^2 {\xi}'{}
-24 {\gamma}'{}+{\sigma}'{}
\right)
\Bigr)
$$
$$
\mbox{}
+{\omega} {x^3}^2 {\phi_3}'{}^2 {\Phi}{}^2 \left(\kappa ^2 \left(16 {\xi}''{}+{\omega}\right)-48 {\gamma}''{}+2 {\sigma}''{}\right)
$$
\begin{equation} 
\mbox{}
+2 {\omega} {x^3} {\phi_3} {\phi_3}'{} {\Phi}{} \left(-24 {\gamma}'{}+{\sigma}'{}+8 \kappa ^2 {\xi}'{}\right)
=0
\label{E01x0x3}
,\end{equation}
$$
{\phi_0}'{} 
\,
\Bigl(
{x^3} {\phi_3}'{} 
\,
\Bigl[
-24 {\gamma}'{}+{\sigma}'{}+8 \kappa ^2 {\xi}'{}
+
{\Phi}{} 
\Bigl(
\kappa ^2 {\omega} 
-24 {\gamma}''{}+{\sigma}''{}
+
8\kappa ^2 {\xi}''{}
\Bigr)
\,\Bigr]
$$
\begin{equation} 
\mbox{}
+{\phi_3} \left(-24 {\gamma}'{}+{\sigma}'{}+8 \kappa ^2 {\xi}'{}\right)
\Bigl)
=0
\label{E03x0x3}
,\end{equation}
$$
{\omega} {\Phi}{}^2 
\left(
{x^3}^2 {\phi_3}'{}^2 \left(\kappa ^2 \left(16 {\xi}''{}+{\omega}\right)-48 {\gamma}''{}+2 {\sigma}''{}\right)
+K \kappa ^2 {\phi_3}^2\right)
$$
$$
\mbox{}
-2{\Phi}{}  {\phi_3} \left({\omega} {x^3} {\phi_3}'{}+K {\phi_3}\right) \left(24 {\gamma}'{}-{\sigma}'{}-8 \kappa ^2 {\xi}'{}\right)
$$
\begin{equation} 
\mbox{}
+144 {\omega} {\phi_3}^2 {\gamma}{}-6 {\omega} {\phi_3}^2 \left({\sigma}{}+8 \kappa ^2 {\xi}{}\right)
+
{2 {\omega} \kappa ^2  {\phi_3}^2}
{\Lambda} 
=0
\label{E22x0x3}
,\end{equation}
$$
{\phi_3}^2 
\Bigl(
2 \kappa ^2 \left(
{\Lambda}
-24 {\xi}{}
\right)
+144 {\gamma}{}-6 {\sigma}{}+K \kappa ^2 {\Phi}{}^2
\Bigr)
-{\omega} \kappa ^2 {x^3}^2 {\phi_3}'{}^2 {\Phi}{}^2
$$
\begin{equation} 
\mbox{}
-6 {x^3} {\phi_3} {\phi_3}'{} {\Phi}{} \left(-24 {\gamma}'{}+{\sigma}'{}+8 \kappa ^2 {\xi}'{}\right)
=0
\label{E22x0x3}
.\end{equation}

To simplify the form of field equations, it is convenient to introduce a new auxiliary function ${Y}({\Phi})$ instead of the function $\sigma(\Phi)$:
\begin{equation}
{Y}({\Phi})=\sigma(\Phi)-24 {\gamma}({\Phi})+8 \kappa ^2 {\xi}({\Phi })
.\end{equation}

Then the expression for the scalar potential takes the form
\begin{equation} 
{V}(\Phi) =
-
 \frac{6
 }{\kappa ^2}
\Bigl(
{Y}({\Phi})
+
12 {\gamma}({\Phi})
+
6\kappa ^2 {\xi}({\Phi})
\Bigr)
+\frac{K {\Phi}^2}{2}
+
{\Lambda}
.\end{equation}

Separating the variables in the equation (\ref{E03x0x3}), we get (due to ${\phi_0}'\ne 0$)
equation on ${\phi_3}(x^3)$ and on the auxiliary function ${Y}(\Phi)$:
\begin{equation}
{x^3}{\phi_3}'
=q\, {\phi_3}
\label{EQphi3Ax0x3}
,\end{equation}
\begin{equation}
q {\Phi} \left({Y}''{}+{\omega} \kappa ^2\right)+(q+1){Y}'{}=0
\label{Yx0x3}
,\end{equation}
where ${q}$ is the separation constant.

The integration of the equation (\ref{Yx0x3}) has a special form at $q=\pm 1/2$, so we consider these cases separately below.


\subsection{Case III.A. 
The scalar field depends on all non-ignored variables, $q\ne \pm 1/2$.}


Using the solutions of the equations (\ref{EQphi3Ax0x3}) and (\ref{Yx0x3}), we obtain the following form of the function $\phi_3(x^3)$, the scalar potential $V({\Phi})$, and functions depending on the scalar field
$\sigma({\Phi} )$ and $\xi({\Phi})$:
\begin{equation}
\phi_3 = p \left({x^3}\right)^{q}
,\qquad
{K} = {\omega}{q} ({q}-3) , \qquad {q}\ne 0,\pm 1/2
\label{phi3Ax0x3}
,\end{equation}
$$
\sigma({\Phi} ) = \frac{-{{a_5}} {q} (2 {q}+1) {\Phi}^{-1/{q}}-\frac{1}{2} {\omega} \kappa ^2 {q} {\Phi}^2}{2 {q}+1}
+24 {\gamma}({\Phi})+\frac{{{{\Lambda}}} \kappa ^2}{3}
$$
\begin{equation}
\mbox{}
-8 \kappa ^2 
\left[
{\Phi}^{-1/{q}} \left(\frac{{a_2} \log ({\Phi})}{2 {q}}+{{a_4}}\right)
+{{a_4}} {\Phi}^{1/q}+\frac{{a_3} {\Phi}^2}{1-4 {q}^2}+\frac{{{{\Lambda}}}}{24}
\right]
\label{sigmaAx0x3}
,\end{equation}
\begin{equation}
\xi({\Phi}) = {\Phi}^{-1/{q}} \left(\frac{{a_2} \log ({\Phi})}{2 {q}}+{{a_4}}\right)
+{{a_4}} {\Phi}^{1/q}+\frac{{a_3} {\Phi}^2}{1-4 {q}^2}+\frac{{{{\Lambda}}}}{24}
\label{xiAx0x3}
,\end{equation}
$$
V({\Phi}) = 6 {\Phi}^{-1/{q}} \left(-\frac{3 {a_2} \log ({\Phi})}{{q}}+\frac{{{a_5}} {q}}{\kappa ^2}-6 {{a_4}}\right)
-36 {{a_4}} {\Phi}^{1/q}
$$
\begin{equation}
\mbox{}
+\frac{
{\omega} {q} 
(q-1)(2q-1)(2q-3)
+72 {a_3}
}{8 {q}^2-2}
\, {\Phi}^2
 -\frac{72 {\gamma}({\Phi})}{\kappa ^2}
 -\frac{5 {{{\Lambda}}}}{2}
\label{VAx0x3}
,\end{equation}
where ${p}$, ${{\Lambda}}$, ${a_2}$, ${a_3}$, ${{a_4}}$ and ${{a_5}}$ are integration constants.
For ${q}= 0$, the case $\Phi=\phi_0(x^0)$ arises, considered earlier and for ${q}=\pm 1/2$, special cases arise, which we will also consider separately below.

One of the functions of the scalar field, namely the function ${\gamma}({\Phi})$ remains arbitrary in the resulting solution, but can be fixed by choosing an additional condition like $\sigma(\Phi)=\mbox{const}$ or conditions $V(\Phi)=\mbox{const}$.

Of the field equations, only one equation remains, which does not turn into an identity, which includes a term responsible for pure radiation (which plays the role of a material source in the gravitational wave model under consideration). From this equation, when interpreting pure radiation as a high-frequency component of the gravitational wave itself, its energy density ${\varepsilon}(x^0,x^3)$ is determined:
$$
{q} {\phi_0}{}^2 
\Bigl[
\kappa ^2 {\Phi}^{\frac{1}{{q}}} \left({\alpha} ({\alpha}-2) ({\omega} {q}+16 (2 {q} {a_3}+{a_3})) {\Phi}^2
-8 (2 {q}+1) {x^0}^2 {L_0}^2 {\varepsilon}{}\right)
$$
$$
\mbox{}
+2 {\alpha} ({\alpha}-2) (2 {q}+1) \left({{a_5}} {q}+8 \kappa ^2 {a_2}\right)
\Bigr]
$$
$$
\mbox{}
-8 {q} {x^0}^2 {\phi_0}{} {\phi_0}''{} \left(-{\omega} \kappa ^2 {q} {\Phi}^{\frac{1}{{q}}+2}+2 {{a_5}} {q}+{{a_5}}\right)
$$
\begin{equation}
\mbox{}
+8 ({q}+1) {x^0}^2 {\phi_0}'{}^2 \left(-{\omega} \kappa ^2 {q} {\Phi}^{\frac{1}{{q}}+2}+2 {{a_5}} {q}+{{a_5}}\right)
=0
\label{varepsilonx0x3}
.\end{equation}

If pure radiation in the equation (\ref{varepsilonx0x3}) is interpreted  as external radiation and is given explicitly, then this equation serves to determine the function ${\phi_0}(x^0)$ in the scalar field ${\Phi}( x^0,x^3)$.

The equation (\ref{varepsilonx0x3}) for the special vacuum case, when there is no pure radiation and ${\varepsilon}=0$, can be written as
(${q}\ne 0, \pm 1/2$):
$$
\kappa ^2 {q}{\Phi}^{2+\frac{1}{{q}}}
\biggl[
{\phi_0}{}^2
{\alpha} 
({\alpha}-2) 
\Bigl({\omega} {q}+16{a_3} (2 {q} +1)\Bigr) 
+
8
{\omega} 
{x^0}^2
\Bigl(
{q} 
{\phi_0}{} {\phi_0}''{} 
-
({q}+1) 
{\phi_0}'{}^2 
\Bigr)
\biggr]
$$
$$
\mbox{}
+
2
{q} {\phi_0}{}^2 
{\alpha} ({\alpha}-2) (2 {q}+1) \left({{a_5}} {q}+8 \kappa ^2 {a_2}\right)
$$
\begin{equation}
\mbox{}
-
8
\left(2  {q}+1\right){{a_5}}
{x^0}^2 
\Bigl(
{q} 
{\phi_0}{} {\phi_0}''{} 
-
({q}+1) 
 {\phi_0}'{}^2 
\Bigr)
=0
.\end{equation}
The term in square brackets due to the functional independence of this term in the equation should vanish separately. Then we obtain the condition for the compatibility of the arising two independent equations:
\begin{equation}
{\alpha} 
({\alpha}-2) 
\biggl[
\Bigl(
3
{\omega} {q}
+
16{a_3} (2 {q} +1)
\Bigr) 
{{a_5}}
+16 {\omega}  \kappa ^2 {a_2}
\biggr]
=0
\label{qneMinus1Div2SovmConsts}
.\end{equation}

The condition (\ref{qneMinus1Div2SovmConsts}) is satisfied if either the gravity model under consideration is conformally flat, i.e. when equality holds
${\alpha} ({\alpha}-2) =0$,
or this condition imposes a connection on the constants ${q}$, ${a_2}$, ${a_3}$ and ${a_5}$ of the following form:
\begin{equation}
{a_2}=
{{a_5}}\,
\frac{
3{\omega} {q}
+
16{a_3} (2 {q} +1)
}{16 {\omega}  \kappa ^2}
\label{SovmConst1}
.\end{equation}

The constant ${a_2}$ vanishes either when ${{a_5}}=0$ or when the relation
\begin{equation}
{a_3}=
-
\frac{
3{\omega} {q}
}{16 (2 {q} +1)}
\quad\to\quad
{a_2}=0
 .\end{equation}
 
Thus, in the case of vacuum, under the condition (\ref{qneMinus1Div2SovmConsts}), the only equation remains on
$\phi_0(x^0)$ of the following form:
\begin{equation}
\beta
{\phi_0}{}^2
+
{x^0}^2
\Bigl(
{\phi_0}{} {\phi_0}''{} 
-
\frac{({q}+1) }{{q}}
{\phi_0}'{}^2 
\Bigr)
=0
\label{phi0Ax0x3}
,\end{equation}
where the auxiliary notation for the constant is used
\begin{equation}
\beta=
\frac{
{\alpha} 
({\alpha}-2) 
\bigl({\omega} {q}+16{a_3} (2 {q} +1)\bigr) 
}{8 {q}
{\omega} }
=\mbox{const}
.\end{equation}
Integrating the equation (\ref{phi0Ax0x3}),
we obtain for the vacuum case a solution for ${\phi_0}(x^0)$ of the following form:
\begin{equation}
{\phi_0}=
\left(
\frac{
{x^0}^{(\nu-1)/2}
}{
{x^0}^{\nu}+{a_0}
}
\right)^{q}
,\qquad
\nu=\sqrt{\left(1+4\beta/{q}\right)}
=\mbox{const}
,\end{equation}
where ${{a_0}}$ is the constant of integration, ${\nu}$ is an auxiliary notation for the constant.

Thus, the scalar field for the vacuum case takes the following general form:
\begin{equation}
\Phi(x^0,x^3)=
p\,
\left(
\frac{
{x^3}
{x^0}^{(\nu-1)/2}
}{
\left({{a_0}}+{x^0}^{\nu}\right) }
\right)^{q}
,\qquad
q\ne \pm 1/2
,\end{equation}
\begin{equation}
\nu=\sqrt{
1+
{\alpha} 
({\alpha}-2) 
\,
\frac{
{\omega} {q}+16{a_3} (2 {q} +1)
}{2 {\omega} {q}^2 }
}
,\end{equation}
where $p$, $q$, $\nu$ and ${{a_0}}$ are constant parameters.

The remaining functions in the obtained exact solution of the field equations of the quadratic gravity theory with a scalar field $\Phi$ are determined by the relations (\ref{sigmaAx0x3})--(\ref{VAx0x3}). Moreover, for the solution, an additional condition (\ref{SovmConst1}) on the constant ${a_2}$ must be satisfied.

Note that the solutions obtained in this section for the scalar field functions included in the Lagrangian of the considered quadratic gravity theory include terms with a functional dependence on the scalar field of the form ${\Phi}^2$, ${\Phi}^n$, ${\Phi}^{-n}$ and ${\Phi}^n \log{{\Phi}}$, where $n$ is an arbitrary constant not equal to $\pm 2$ ($n=-1 /q$, $q\ne\pm 1/2$).
Special cases of $q=\pm 1/2$ will be considered below.


\subsection{Case III.B. 
The scalar field depends on all non-ignored variables, $q=-1/2$.}


Let us now consider a special case when the constant ${q} = -1/2$ in the equation (\ref{Yx0x3}) and the functions $\sigma({\Phi})$ and $\phi_3(x^3)$ during integration of the corresponding equation take a special form
(the function ${\phi_0}(x^0)$ is still undefined):
\begin{equation} 
{q} = 
-1/2
.\end{equation}
\begin{equation} 
\Phi(x^0,x^3)=\phi_0(x^0)\phi_3(x^3)
.\end{equation}
\begin{equation} 
\phi_3 = \frac{p}{\sqrt{{x^3}}}
,\qquad
p-\mbox{const}
.\end{equation}
\begin{equation} 
\sigma({\Phi}) = \frac{{{b_2}}}{2}  {\Phi}^2+\frac{{\omega} \kappa ^2}{4}  {\Phi}^2
-\frac{{\omega} \kappa ^2}{2}  {\Phi}^2 \log ({\Phi})+24 {\gamma}({\Phi})-8 \kappa ^2 {\xi}({\Phi})+\frac{{{{\Lambda}}} \kappa ^2}{3}
.\end{equation}
\begin{equation} 
V({\Phi}) = {\Phi}^2 \left(3 {\omega} \log ({\Phi})-\frac{3 {{b_2}}}{\kappa ^2}-\frac{5 {\omega}}{8}\right)-\frac{72 {\gamma}({\Phi})}{\kappa ^2}-36 {\xi}({\Phi})-{{{\Lambda}}}
,\end{equation}
where
${{\Lambda}}$ and
${b_2}$ are the integration constants of the field equations.
The functions ${\gamma}({\Phi})$ and ${\xi}({\Phi})$ are still undefined.

As a result, only one equation from the field equations does not turn into an identity:
$$
48\kappa ^2{L_0}{}^2 {\varepsilon}(x^0,x^3) =
\frac{
6{\omega} \kappa ^2 {\Phi}^2 \log ({\Phi})
}{{x^0}^2 {\phi_0}{}^2}
\biggl[
{\alpha}
({\alpha}-2)  \, 
{\phi_0}{}^2 
+
8{x^0}^2 
\left(
{\phi_0}{} {\phi_0}''{}
+
{\phi_0}'{}^2 
\right)
\biggr]
$$
$$
\mbox{}
-
\frac{
3 {\Phi}^2 
}{{x^0}^2 {\phi_0}{}^2}
\,
\biggl[
\,
{\alpha} ({\alpha}-2) {\phi_0}{}^2 \left(
8 \kappa ^2 {\xi}''{}+2 {{b_2}}+{\omega} \kappa ^2\right)
+
16 {{b_2}} {x^0}^2 
\left(
{\phi_0}{} {\phi_0}''{} 
+
{\phi_0}'{}^2 
\right)
\,\biggr]
$$
\begin{equation} 
-
4 \kappa ^2 
\,
\frac{{\alpha}({\alpha}-2)  ({{{\Lambda}}}-24 {\xi}({\Phi}))}{{x^0}^2}
-
\frac{24 {\alpha} ({\alpha}-2) \kappa ^2 {\Phi} {\xi}'{}}{{x^0}^2}
\label{E00qEqMinus1Div2}
.\end{equation}

The equation (\ref{E00qEqMinus1Div2}) defines pure radiation in the problem under consideration, namely, it gives an expression for the energy density of pure radiation ${\varepsilon}(x^0,x^3)$.
In this case, the functions $\gamma(\Phi)$, ${\xi}(\Phi)$ and the function $\phi_0(x^0)$ remain arbitrary (undefined).

Pure radiation can be interpreted as a high-frequency, high-energy component of the gravitational wave itself, propagating along the variable $x^1$ of the privileged coordinate system $L^{\beta}=\{0,L^1,0,0\}$.

In another interpretation, when pure radiation is external and its functional form is given, then the equation (\ref{E00qEqMinus1Div2}) defines the function $\phi_0(x^0)$ and the function ${\xi}({\Phi})$. The function $\gamma(\Phi)$ remains arbitrary in this case.

In the special case of vacuum, when the energy density of pure radiation vanishes ${\varepsilon}=0$, we obtain an additional relation for the function $\phi_0(x^0)$ and the function ${\xi}(\Phi)$ of the following form :
$$
3 {\Phi}^2 
\,
\biggl[
\,
{\alpha} ({\alpha}-2) 
\left(
2 {{b_2}}+{\omega} \kappa ^2\right)
+
16 {{b_2}} \,
\frac{
{x^0}^2 
}{{\phi_0}{}^2}
\left(
{\phi_0}{} {\phi_0}''{} 
+
{\phi_0}'{}^2 
\right)
\,\biggr]
$$
$$
-
6{\omega} \kappa ^2 {\Phi}^2 \log ({\Phi})
\biggl[
{\alpha}
({\alpha}-2)  
+
8\,
\frac{
{x^0}^2 
}{ {\phi_0}{}^2}
\left(
{\phi_0}{} {\phi_0}''{}
+
{\phi_0}'{}^2 
\right)
\biggr]
$$
\begin{equation}
\mbox{}
+
4 \kappa ^2
{\alpha} ({\alpha}-2) 
\,
\Bigl(
6 {\Phi}^2 
{\xi}''{}
+
{{{\Lambda}}}-24 {\xi}({\Phi})
+
6
{\Phi} {\xi}'{}
\Bigr)
=0
\label{E00qEqMinus1Div2Vacuum}
.\end{equation}

The relation (\ref{E00qEqMinus1Div2Vacuum}) implies two independent equations for the function
${\xi}({\Phi})$ and the function ${\phi_0}(x^0)$.

One of the equations arising from (\ref{E00qEqMinus1Div2Vacuum}) is an equation for the function ${\xi}({\Phi})$ of the following form:
\begin{equation}
6 
{\Phi} 
\left(
{\Phi}
{\xi}''{}
+
{\xi}'{}
\right)
-24 {\xi}({\Phi})
+
{{{\Lambda}}}
=
{b_3}\Phi^2+{b_4}\Phi^2\log\Phi
,\qquad
{b_3},{b_4} - \mbox{const}
\label{xiVacuumx0x3}
,\end{equation}
where ${b_3}$ and ${b_4}$ are constants in the expansion of expressions in independent functions of the scalar field.

Integrating the equation (\ref{xiVacuumx0x3}), we obtain the function ${\xi}({\Phi})$ in the form:
$$
{\xi}({\Phi})=
{b_5} \Phi^2+\frac{{b_6}}{\Phi^2}
+\mbox{}
$$
\begin{equation}
\mbox{}
+
\frac{1}{384} 
\Bigl[
16 {{{\Lambda}}}+({b_4}-4 {b_3})\,\Phi^2 +4(4 {b_3}-{b_4})\,  \Phi^2 \log (\Phi)+8 {b_4} \Phi^2 \log^2(\Phi)
\Bigr]
,\end{equation}
where ${b_5}$ and ${b_6}$ are constants of integration.

The second relation arising from the equation (\ref{E00qEqMinus1Div2Vacuum}) gives conditions on the function $\phi_0(x^0)$:
$$
3 {\Phi}^2 
\,
\biggl[
\,
{\alpha} ({\alpha}-2) 
\left(
{\omega} \kappa ^2
+
\frac{4\kappa ^2}{3}
{b_3}
+
2 {{b_2}}
\right)
+
16 {{b_2}} \,
\frac{
{x^0}^2 
}{{\phi_0}{}^2}
\left(
{\phi_0}{} {\phi_0}''{} 
+
{\phi_0}'{}^2 
\right)
\,\biggr]
$$
\begin{equation}
\mbox{}
-
6{\omega} \kappa ^2 {\Phi}^2 \log ({\Phi})
\biggl[
{\alpha} ({\alpha}-2) 
\left(1
-
\frac{2{b_4}}{3\,\omega}
\right)
+
8\,
\frac{
{x^0}^2 
}{ {\phi_0}{}^2}
\left(
{\phi_0}{} {\phi_0}''{}
+
{\phi_0}'{}^2 
\right)
\biggr]
=0
.\end{equation}
Moreover, due to functional independence, the expressions in square brackets must vanish individually. Then we get the condition of their compatibility in the form:
\begin{equation}
{\alpha} ({\alpha}-2)
\biggl[
{\omega} \kappa ^2
+
\frac{4\kappa ^2}{3}{b_3}
+
{{b_2}}
\frac{4{b_4}}{3\,\omega}
\biggr]
=0
\label{SovmConsts}
.\end{equation}
This condition for the case of a conformally flat spacetime, when ${\alpha} ({\alpha}-2) =0$, is satisfied identically, or
from this relation one can find one of the constants ${b_3}$, ${b_4}$ or ${{b_2}}$.

When the condition
(\ref{SovmConsts}) remains the only equation on ${\phi_0}(x^0)$:
\begin{equation}
{\alpha} ({\alpha}-2) 
\left(1
-
\frac{2{b_4}}{3\,\omega}
\right)
+
8\,
\frac{
{x^0}^2 
}{ {\phi_0}{}^2}
\left(
{\phi_0}{} {\phi_0}''{}
+
{\phi_0}'{}^2 
\right)=0
\label{E00qEqMinus1Div2VacuumPhi0}
.\end{equation}

Integrating the equation (\ref{E00qEqMinus1Div2VacuumPhi0}), we get the solution in the form:
\begin{equation}
\phi_0(x^0)=
{x^0}^{(1-\beta)/4} \sqrt{{x^0}^{\beta}+{b_0}}
,
\qquad
{b_0}
 - \mbox{const}
,\end{equation}
$$
\beta=
\sqrt{
1-
{\alpha} ({\alpha}-2) 
\left(1
-
\frac{2{b_4}}{3\,\omega}
\right)
}
=
\mbox{const}
,$$
where ${b_0}$ is the constant of integration, $\beta$ is an auxiliary notation for the constant.

The final vacuum solution of the field equations in the special case when $q=-1/2$ takes the following form:
\begin{equation}
\Phi(x^0,x^3)=\phi_0(x^0)\,\phi_3(x^3)=
p\,
{x^0}^{(1-\beta)/4}
\sqrt{
 \frac{{{x^0}^{\beta}+{b_0}}}{{x^3}}
 }
,\end{equation}
$$
{\xi}({\Phi})=
{b_5} \Phi^2+\frac{{b_6}}{\Phi^2}
+\mbox{}
$$
\begin{equation}
\mbox{}
+
\frac{1}{384} 
\Bigl[
16 {{{\Lambda}}}+({b_4}-4 {b_3})\,\Phi^2 +4(4 {b_3}-{b_4})\,  \Phi^2 \log (\Phi)+8 {b_4} \Phi^2 \log^2(\Phi)
\Bigr]
,\end{equation}
\begin{equation}
\sigma({\Phi}) = \frac{1}{2} {{b_2}} {\Phi}^2+\frac{1}{4} {\omega} \kappa ^2 {\Phi}^2-\frac{1}{2} {\omega} \kappa ^2 {\Phi}^2 \log ({\Phi})+24 {\gamma}({\Phi})-8 \kappa ^2 {\xi}({\Phi})+\frac{{{{\Lambda}}} \kappa ^2}{3}
,\end{equation}
\begin{equation}
V({\Phi}) = {\Phi}^2 \left(3 {\omega} \log ({\Phi})-\frac{3 {{b_2}}}{\kappa ^2}-\frac{5 {\omega}}{8}\right)
-{{{\Lambda}}}
-\frac{72 {\gamma}({\Phi})}{\kappa ^2}-36 {\xi}({\Phi})
.\end{equation}
The function ${\gamma}({\Phi})$ remains arbitrary in this solution.
The solution also contains a set of constants and
a compatibility condition (\ref{SovmConsts}) linking the constants ${b_3}$, ${b_4}$ and ${{b_2}}$.

The remaining arbitrary function ${\gamma}({\Phi})$ can be defined by choosing a condition on $\sigma(\Phi)$ or on the scalar potential $V({\Phi})$. For example, $\sigma(\Phi)=\mbox{const}$ (Jordan frame), or $V({\Phi}) =\mbox{const}$ (cosmological constant).

For example, when choosing the additional condition $\sigma(\Phi)=1$, the function ${\gamma}({\Phi})$ becomes
\begin{equation}
{\gamma}({\Phi})
=
\frac{1}{3}
\,
\left[
\frac{1}{8}
-\frac{{{{\Lambda}}} \kappa ^2}{26}
+
\kappa ^2 {\xi}({\Phi})
- \frac{{\Phi}^2}{16} 
\left(
{{b_2}} 
-\frac{1}{2} {\omega} \kappa ^2 
\right)
+\frac{{\omega} \kappa ^2}{32}  
\,
{\Phi}^2 \log ({\Phi})
\right]
.\end{equation}
And when choosing another additional condition $V({\Phi}) =0$, the function ${\gamma}({\Phi})$ takes the following form
\begin{equation}
{\gamma}({\Phi})
= 
\frac{{\kappa ^2}}{72}
\,
\left[
-{{{\Lambda}}}
-36 {\xi}({\Phi})
+
{\Phi}^2 \left(3 {\omega} \log ({\Phi})-\frac{3 {{b_2}}}{\kappa ^2}-\frac{5 {\omega}}{8}\right)
\right]
.\end{equation}

Thus, the functions of the scalar field entering into the Lagrangian of the quadratic theory of gravity
in the considered model of a gravitational wave in a special case $q=-1/2$ include
terms containing the scalar field in the form ${\Phi}^{-2}$, ${\Phi}^2$, 
${\Phi}^2\log ({\Phi})$ and $\Phi^2 \log^2(\Phi)$.


\subsection{Case III.C. 
The scalar field depends on all non-ignored variables, $q=1/2$.}


Let us now consider a special case of solving field equations for the value of the parameter $q=1/2$, when for a scalar field we have:
\begin{equation}
\phi_3(x^3)=p\sqrt{x^3}
.\end{equation}

The scalar field takes the form
\begin{equation}
\Phi(x^0,x^3)=\phi_0(x^0) \phi_3(x^3)=p\sqrt{x^3} \phi_0(x^0) 
.\end{equation}

The equation (\ref{Yx0x3}) with $q=1/2$ has the following solution
\begin{equation}
Y({\Phi})
=
\sigma(\Phi)-24 {\gamma}({\Phi})+8 \kappa ^2 {\xi}({\Phi})
=
-\frac {\kappa^2\omega}{8}\, {\Phi}^2 +\frac{{c_2}}{{\Phi}^2}+{{c_1}}
,\end{equation}
where ${{c_2}}$ and ${{c_1}}$ are the expansion constants of the function $Y({\Phi})$ in terms of the independent functions of the scalar field in the equation.

Substituting this solution into the remaining field equations, we obtain
\begin{equation}
{{c_1}}=\frac{\kappa^2\Lambda}{3}
.\end{equation}

The scalar field functions take the form:
\begin{equation}
\sigma(\phi ) = \frac{{{c_2}}}{{\Phi}{}^2}-\frac{1}{8} {\omega} \kappa ^2 {\Phi}{}^2
+24 {\gamma}({\Phi}{})-8 \kappa ^2 {\xi}({\Phi}{})+\frac{{\Lambda} \kappa ^2}{3}
,\end{equation}
\begin{equation}
V(\phi ) = -\frac{6 {{c_2}}}{\kappa ^2 {\Phi}{}^2}+\frac{1}{8} {\omega} {\Phi}{}^2
-\frac{72 {\gamma}({\Phi}{})}{\kappa ^2}-36 {\xi}({\Phi}{})-{\Lambda}
,\end{equation}
where the functions $\gamma(\phi )$ and ${\xi}({\Phi}{})$ remain arbitrary.

In this case, of all the field equations, only one equation remains, which does not turn into an identity, of the following form:
$$
{\phi_0}{}^2 
\Bigl[
\kappa ^2 {\Phi}{}^2 
\,
\Bigl(
3 {\alpha} ({\alpha}-2) \left({\Phi}{}^2 \left({\omega}-16 {\xi}''{}\right)-16 {\Phi}{} {\xi}'{}
+64 {\xi}({\Phi}{})\right)
$$
$$
\mbox{}
-8 \left(12 {x^0}^2 {L_0}{}^2 {\varepsilon}{}+{\alpha} ({\alpha}-2) {\Lambda}\right)
\Bigr)
-24 {{c_2}} {\alpha} ({\alpha}-2)
\Bigr]
$$
\begin{equation}
\mbox{}
+24 {x^0}^2 {\phi_0}{} {\phi_0}''{} \left({\omega} \kappa ^2 {\Phi}{}^4+8 {{c_2}}\right)-72 {x^0}^2 {\phi_0}'{}^2 \left({\omega} \kappa ^2 {\Phi}{}^4+8 {{c_2}}
\right)
=0
\label{varepsilonx0x3Plus1Div2}
.\end{equation}

This equation is related to the definition of the pure radiation energy density function
${\varepsilon}(x^0,x^3)$. The equation (\ref{varepsilonx0x3Plus1Div2}) contains as unknown quantities, in addition to ${\varepsilon}$, the function $\phi_0(x^0)$ entering the scalar field ${\Phi}$ and the unknown function of the scalar field ${\xi}({\Phi}{})$.

If we assume that the type of external pure radiation is given, then the relation (\ref{varepsilonx0x3Plus1Div2}) is an equation for $\phi_0(x^0)$. Moreover, in this case, due to the presence of functions in the equation (\ref{varepsilonx0x3Plus1Div2}) that depend on both $x^0$ and $x^3$, it is necessary to solve this equation as a functional equation.

If pure radiation is interpreted as the high-frequency part of the gravitational wave itself on the basis of the considered metric, then the relation (\ref{varepsilonx0x3Plus1Div2}) determines the energy density of pure gravitational radiation ${\varepsilon}(x^0,x^3)$.

Let us consider the vacuum case, when the energy density of pure radiation vanishes, i.e.
$\varepsilon=0$.

Then from the equation (\ref{varepsilonx0x3Plus1Div2}), considered as a functional equation, it follows that either ${\alpha} ({\alpha}-2)=0$, or the following condition is satisfied:
\begin{equation}
{\Phi}{}^2 
\left[
{\Phi}{}^2 \left({\omega}-16 {\xi}''{}\right)-16 {\Phi}{} {\xi}'{}+64 {\xi}({\Phi}{})
\right]
=
{{c_3}}{\Phi}{}^4+{{c_4}}{\Phi}{}^2+{{c_5}}
\label{E00x0x3qPlus1Div2}
,\end{equation}
where ${{c_3}}$, ${{c_4}}$ and ${{c_5}}$ are the constant parameters of the expansion of the expression on the left into independent functions of the scalar field included in the 
equation (\ref{varepsilonx0x3Plus1Div2}).

Integrating the equation (\ref{E00x0x3qPlus1Div2}), we get:
\begin{equation}
\xi ({\Phi})
= 
\frac{
4 \left({\Phi}^4 (\omega-{{c_3}})+{{c_5}}\right)\log ({\Phi}) 
+
{\Phi}^4
\left(
{{c_3}}
-\omega
+256{c_7} 
\right)
+4 {{c_4}} {\Phi}^2
+{{c_5}}
+256{c_6}
}{256 {\Phi}^2}
,\end{equation}
where ${c_6}$ and ${c_7}$ are integration constants.

Substituting the resulting expression for $\xi ({\Phi})$ into the equation (\ref{varepsilonx0x3Plus1Div2}) and expanding it into a series in powers of the scalar field, we obtain the relation of the following form:
$$
0=
{\Phi}{}^4
\left[
{\alpha} ({\alpha}-2) 3\kappa ^2 {{c_3}}
+
\frac{
24{x^0}^2  }{{\phi_0}{}^2 }
{\omega} \kappa ^2
\Bigl(
{\phi_0}{} {\phi_0}''{}
-3  {\phi_0}'{}^2 
\Bigr)
\right]
$$
$$
\mbox{}
+
{\Phi}{}^2
{\alpha} ({\alpha}-2) 
\kappa ^2
\left(
3 {{c_4}}
-
8 {\Lambda}
\right)
$$
\begin{equation}
\mbox{}
+
{\alpha} ({\alpha}-2) 
\left(
3\kappa ^2 {{c_5}}
-24 {{c_2}} 
\right)
+
\frac{
24{x^0}^2  }{{\phi_0}{}^2 }
8 {{c_2}}
\Bigl(
{\phi_0}{} {\phi_0}''{} 
-3  {\phi_0}'{}^2 
\Bigr)
\label{E00V2x0x3qPlus1Div2}
.\end{equation}

It immediately follows from this equation that
\begin{equation}
{{c_4}}=\frac{8\Lambda}{3}
.\end{equation}
The remaining two independent terms in the equation (\ref{E00V2x0x3qPlus1Div2})
for ${\alpha} ({\alpha}-2) \ne 0$
joint on the condition that:
\begin{equation}
{{c_5}}=\frac{8{{c_2}}({{c_3}}+{\omega})}{{\omega}\kappa ^2}
.\end{equation}
There remains only one field equation for the function ${\phi_0}(x^0)$:
\begin{equation}
{\alpha} ({\alpha}-2)  {{c_3}}
{{\phi_0}{}^2 }
+
8
{\omega} 
{x^0}^2  
\Bigl(
{\phi_0}{} {\phi_0}''{}
-3  {\phi_0}'{}^2 
\Bigr)
=0
.\end{equation}
Integrating this equation
we get the following solution
for function $\phi_0$:
\begin{equation}
\phi_0 (x^0)= 
\frac{
{x^0}^{\left(\beta-1\right)/4}
}{\sqrt{{x^0}^{\beta}+
{{c_0}}
}}
,\qquad
\beta=\sqrt{1+\frac{ {\alpha} ({\alpha}-2) 
}{{\omega}}
\,
{{c_3}}
}
=\mbox{const}
,\end{equation}
where ${{c_0}}$ and ${{c_3}}$ are independent constant parameters, $\beta$ is an auxiliary notation for a combination of constant parameters.

Finally, the scalar field in vacuum in the case of $q=1/2$ takes the form:
\begin{equation}
{\Phi}(x^0,x^3)=
{p}\,
\sqrt{{x^3}\,\frac{{x^0}^{\left(\beta-1\right)/2}}{{x^0}^{\beta}+{{c_0}}}}
.\end{equation}

Also, from the equation (\ref{varepsilonx0x3Plus1Div2}), the explicit form of the function $\xi(\phi ) $ is determined:
$$
\xi({\Phi}) = 
\frac{
1}{256 {\Phi}^2}
\biggl[
4 \log ({\Phi}) 
\left(
({\omega}-{{c_3}}) {\Phi}^4+\frac{8 {{c_2}} ({{c_3}}+{\omega})}{{\omega} \kappa ^2}
\right)
$$
\begin{equation} 
\mbox{}
+
{\Phi}^4
\left(
{{c_3}} 
-{\omega} 
+256{c_7} 
\right)
+\frac{32}{3} {\Lambda} {\Phi}^2
+\frac{8 {{c_2}} ({{c_3}}+{\omega})}{{\omega} \kappa ^2}+256{c_6}
\biggr]
.\end{equation} 

The remaining functions of the scalar field take the following form:
$$
\sigma({\Phi}) =
-
\frac{
1}{32 {\omega} {\Phi}^2}
\biggl[
4 \log ({\Phi}) 
\Bigl(
8 {{c_2}} ({{c_3}}+{\omega})+{\omega} \kappa ^2 {\Phi}^4 ({\omega}-{{c_3}})
\Bigr)
+8 {{c_2}} ({{c_3}}-3
   {\omega})
$$
\begin{equation} 
\mbox{}
+{\omega} \kappa ^2 \left({\Phi}^4 ({{c_3}}+3 {\omega}+256{c_7})+256{c_6}\right)-768 {\omega} {\Phi}^2
   {\gamma}({\Phi})
\biggr]
,\end{equation} 
$$
V({\Phi}) = -3\,\frac{
12 {{c_2}} ({{c_3}}+{\omega}) \log ({\Phi})+3 {{c_2}} {{c_3}}
+19 {{c_2}} {\omega}
+96 {\omega} \kappa ^2{c_6}
}{8 {\omega} \kappa ^2 {\Phi}^2}
$$
\begin{equation} 
\mbox{}
+\frac{{\Phi}^2}{64} 
\Bigl(
36 ({{c_3}}-{\omega}) \log ({\Phi})-9 {{c_3}}+17 {\omega}-2304{c_7}
\Bigr)
-\frac{5 {\Lambda}}{2}
-\frac{72 {\gamma}({\Phi})}{\kappa ^2}
.\end{equation}
In this case, one of the functions of the scalar field, ${\gamma}({\Phi})$, remains arbitrary in the resulting vacuum solution.

Note that there are two relations for the three functions of the scalar field $\sigma({\Phi})$, $\gamma({\Phi})$ and the scalar potential $V({\Phi})$. One of these functions remains arbitrary in the solution.

If $\sigma$ is fixed, i.e. set the type of this coefficient
for a term linear in the scalar curvature ${R}$ in the Lagrangian,
then we obtain the following relations for the remaining functions of the scalar field:
$$
\gamma(\Phi)
= 
\frac{
1}{768 {\omega} {\Phi}^2}
\,
\biggl[
4 \log ({\Phi}) 
\Bigl(
8 {{c_2}} ({{c_3}}+{\omega})+{\omega} \kappa ^2 {\Phi}^4  ({\omega}-{{c_3}})
\Bigr)
+8 {{c_2}} ({{c_3}}-3 {\omega})
$$
\begin{equation} 
\mbox{}
+{\omega} 
\Bigl(
{{c_3}} \kappa ^2 {\Phi}^4
+3 {\omega} \kappa ^2  {\Phi}^4
+32 \left(8 \kappa ^2 {\Phi}^4{c_7}
+8 \kappa ^2{c_6}+{\Phi}^2 \sigma 
\right)
\Bigr)
\biggr]
,\end{equation} 
$$
V(\Phi)=
-
\frac{
1}{64 {\omega} \kappa ^2 {\Phi}^2}
\biggl[
60 \log ({\Phi}) 
\Bigl(
8 {{c_2}} ({{c_3}}+{\omega})+{\omega} \kappa ^2 {\Phi}^4 ({\omega}
-{{c_3}})
\Bigr)
+24 {{c_2}} (5 {{c_3}}+13 {\omega})
$$
\begin{equation} 
\mbox{}
+
{\omega} 
\Bigl(
\kappa ^2
{\Phi}^4
\left(
{\omega} 
+
15 {{c_3}}  
+3840  {c_7}
\right)
+160 {\Lambda} \kappa ^2 {\Phi}^2+192 {\Phi}^2 \sigma
+3840 \kappa ^2{c_6}
\Bigr)
\biggr]
.\end{equation} 

If, instead of fixing $\sigma$, we specify the form of the potential of the scalar field $V({\Phi})$,
then in this case we obtain the following relations for the remaining functions of the scalar field:
$$
\gamma(\Phi)=
\frac{
1}{4608 {\omega}{\Phi}^2}
\biggl[
-36 \log ({\Phi}) 
\Bigl(
8 {{c_2}} ({{c_3}}+{\omega})+{\omega} \kappa ^2 {\Phi}^4
   ({\omega}-{{c_3}})
\Bigr)
   -24 {{c_2}} (3 {{c_3}}+19 {\omega})
$$
\begin{equation} 
\mbox{}
+{\omega} \kappa ^2 \left(
 {\Phi}^4
\Bigl(
17 {\omega}
-9 {{c_3}} 
-2304 {c_7}
\Bigr)
-160 {\Lambda}   {\Phi}^2-64 {\Phi}^2 {V}-2304{c_6}\right)
\biggr]
,
\end{equation} 
$$
\sigma(\Phi)=
 -
\frac{
1}{192 {\omega} {\Phi}^2}
\biggl[
 60 \log ({\Phi}) 
\Bigl(
8 {{c_2}} ({{c_3}}+{\omega})+{\omega} \kappa ^2 {\Phi}^4
   ({\omega}-{{c_3}})
\Bigr)
+24 {{c_2}} (5 {{c_3}}+13 {\omega})
$$
\begin{equation} 
\mbox{}
+{\omega} \kappa ^2 
\left(
{\Phi}^4
\Bigl(
15 {{c_3}} 
+{\omega}
+3840{c_7}
\Bigr)
+160 {\Lambda} {\Phi}^2
+64 {\Phi}^2 {V}+3840{c_6}
\right)
\biggr]
.\end{equation} 

Thus, the functions of the scalar field included in the Lagrangian of the considered quadratic theory of gravity in the special case $q=1/2$ include terms that depend on the scalar field as follows
${\Phi}^{-2}$, ${\Phi}^2$,
${\Phi}^{-2}\log ({\Phi})$, ${\Phi}^2\log ({\Phi})$.

Obtaining an explicit functional form for the scalar field functions in action for the quadratic theory of gravity (\ref{Lagrangian}) is a significant result of the considered model.

\section{Radiation Propagation}

To find out the trajectories of radiation propagation in the models under consideration (the trajectories of light rays), we can use the Hamilton-Jacobi formalism.

Propagation of light rays
described by the eikonal equation
\begin{equation}
g^{\mu\nu}\frac{\partial\Psi}{\partial x^\mu}\frac{\partial\Psi}{\partial x^\nu}=0
\label{EqEikonal}
,\end{equation}
where ${\Psi}$ is the eikonal function.

For type II Shapovalov wave spacetimes with two ignored variables $x^1$ and $x^2$, to which the considered metric belongs, the solution of the eikonal equation can be written in the ''separated'' form:
\begin{equation}
{\Psi}={p_1} x^1+{p_2} x^2+\psi_0(x^0)+\psi_3(x^3)
+\mbox{const}
,\end{equation}
where ${p_1}$ and ${p_2}$ are independent constant parameters.

Separating the variables in the eikonal equation (\ref{EqEikonal}), we get
\begin{equation}
{\psi_3}'(x^3)=\mbox{const}={p_3}
,\end{equation}
\begin{equation}
{p_1}
{\psi_0}'(x^0)
=
-\frac{1}{2}
\,
\left({p_3}+{p_2}^2 \,{x^0}^{-\alpha}\right)
,\end{equation}
where ${p_3}$ is an additional independent constant partition parameter.
Note that when ${p_1}=0$, then the other constants also vanish, i.e. ${p_2}={p_3}=0$, and the considered trajectory degenerates. 

Therefore, further consider the case ${p_1}\ne 0$, then we have
\begin{equation}
{\psi_0}(x^0)
=
-\frac{1}{2p_1}
\,
\left({p_3}x^0 
+{p_2}^2 \,\int{{x^0}^{-\alpha}\,dx^0}
\right)
,\qquad
{\psi_3}(x^3)={p_3}x^3
.\end{equation}

For $\alpha=1$ the integral for ${\psi_0}$ has a special form, and this case should be considered separately:
\begin{equation}
\alpha=1
\quad
\to
\quad
{\psi_0}(x^0)
=
-\frac{1}{2p_1}
\,
\left({p_3}x^0 
+{p_2}^2 \,\log{x^0}
\right)
.\end{equation}
\begin{equation}
\alpha\ne 1
\quad
\to
\quad
{\psi_0}(x^0)
=
-\frac{1}{2p_1}
\,
\left({p_3}x^0 
+\frac{{p_2}^2}{1-\alpha}
\,
{x^0}^{1-\alpha}
\right)
.\end{equation}

In accordance with the Hamilton-Jacobi formalism, the equations for the trajectory of a light beam can be found from the eikonal function ${\Psi}$ from the relations
\begin{equation}
\frac{\partial \Psi}{\partial {p_k} }
=
{q_k}
,\qquad
{q_k}=\mbox{const}
,\qquad
k=1,2,3
\label{TrajectorEqs}
,\end{equation}
where ${q_k}$ are new independent constant parameters of the light beam trajectory equations determined by the initial conditions.

\subsection{Radiation propagation. Metric at $\alpha=1$.}

For $\alpha=1$, the light beam trajectory equations (\ref{TrajectorEqs}) take the form
\begin{equation}
x^1-q_1
+
\frac{1}{2{p_1}^2}
\,
\left({p_3}x^0 
+{p_2}^2 \,\log{x^0}
\right)
=0
,\end{equation}
\begin{equation}
x^2-q_2
-\frac{p_2}{p_1}\,\log{x^0}
=0
,\end{equation}
\begin{equation}
x^3-q_3
-\frac{x^0}{2{p_1}}
=0
,\end{equation}
where by shifting the origin of the variables $x^k$ we can set the constants $q_k$ equal to zero, the constants $p_k$ are determined by the initial conditions.

Choosing the wave variable $x^0$ as the parameter ${\tau}$ (${\tau}>0$) on the trajectories of the light beam, we obtain the parametric form of the trajectory equations in the coordinates $x^k(\tau)$:
\begin{equation}
x^1({\tau})=
-
\frac{1}{2{p_1}^2}
\,
\left({p_3}{\tau}
+{p_2}^2 \,\log{{\tau}}
\right)
,\end{equation}
\begin{equation}
x^2({\tau})=
\frac{p_2}{p_1}\,\log{{\tau}}
,\end{equation}
\begin{equation}
x^3({\tau})=
\frac{{\tau}}{2{p_1}}
.\end{equation}

For the coordinates $x^k$ on the trajectories of the light beam, the following relations hold:
\begin{equation}
2{p_1}^2\, x^1+
2{p_1}{p_3}\, x^3
+{p_2}^2 \,\log{x^3}
=
-{p_2}^2 \,\log{\left(2{p_1}\right)}
=
\mbox{const}
,\end{equation}
\begin{equation}
{p_1}x^2
-
{p_2}\log{x^3}
=
{p_2}\log{\left(2{p_1}\right)}
=
\mbox{const}
.\end{equation}

These relations give a connection between the coordinates $x^k$ on the trajectories of light rays in the considered models of a gravitational wave in privileged coordinate systems with a wave variable $x^0$.

\subsection{Propagation of radiation. Metric at $\alpha\ne 1$.}

For $\alpha\ne 1$, the light beam trajectory equations (\ref{TrajectorEqs}) take the form
\begin{equation}
x^1-q_1
+
\frac{1}{2{p_1}^2}
\,
\left({p_3}x^0 
+\frac{{p_2}^2}{1-\alpha} \,{x^0}^{1-\alpha}
\right)
=0
,\end{equation}
\begin{equation}
x^2-q_2
-\frac{p_2}{p_1(1-\alpha)}\,{x^0}^{1-\alpha}
=0
,\end{equation}
\begin{equation}
x^3-q_3
-\frac{x^0}{2{p_1}}
=0
,\end{equation}
where by shifting the origin of the variables $x^k$ we can set the constants $q_k$ equal to zero, ${p_1}$, ${p_2}$ and ${p_3}$ are independent constant parameters determined by the initial conditions.

Choosing the wave variable $x^0$ as the parameter ${\tau}$ (${\tau}>0$) on the trajectories of the light beam, we obtain the parametric form of the trajectory equations in the coordinates $x^k({\tau})$ :
\begin{equation}
x^1({\tau})=
-
\frac{1}{2{p_1}^2}
\,
\left({p_3}{\tau}
+\frac{{p_2}^2}{1-\alpha} \,{\tau}^{(1-\alpha)}
\right)
,\end{equation}
\begin{equation}
x^2({\tau})
=
\frac{p_2}{p_1(1-\alpha)}\,{\tau}^{(1-\alpha)}
,\end{equation}
\begin{equation}
x^3({\tau})
=
\frac{\tau}{2{p_1}}
,\end{equation}

For the coordinates $x^k$ on the trajectories of the light beam, the relations
\begin{equation}
2{p_1}^2 x^1+
2{p_1}{p_3}x^3
+\frac{{p_2}^2}{1-\alpha} \,{\left(2{p_1}x^3\right)}^{(1-\alpha)}
=0
,\end{equation}
\begin{equation}
x^2
=
\frac{p_2}{p_1(1-\alpha)}\,{\left(2{p_1}x^3\right)}^{(1-\alpha)}
,\end{equation}
where $\alpha$ is a constant parameter of the gravitational-wave model, $p_k$ are constants determined by the initial conditions.

Thus, we have obtained an explicit form of the light beam trajectories in the considered models 
 in a privileged coordinate system.


\section{Conclusion}

Exact models of the primordial gravitational wave in the Bianchi type III universe are obtained on the basis of the quadratic theory of gravity with a scalar field and pure radiation. The exact solutions of the field equations are found, including the solutions of the scalar equation of the theory, and the explicit form of the scalar field functions entering into the Lagrangian of the theory under consideration is obtained. Relationships are obtained for the characteristics of pure radiation, which is interpreted either as a high-frequency part of a gravitational wave, or as a source of external radiation. For the model of the primordial  gravitational wave under consideration, exact solutions of the eikonal equation for the propagation of radiation are found and the exact form of the trajectories of light rays is obtained.

The exact models of the primordial gravitational wave constructed in this work give rare examples of this kind of exact models. For complex modified theories of gravity with a scalar field that describe the early stages of the universe, obtaining such exact models provides a basis for analyzing the properties of primordial gravitational waves  in general and for complex perturbative models of gravitational waves in particular.


%

\authorcontributions{Conceptualization, K.O.; methodology, K.O. and I.K.; validation, I.K., E.O. and K.O.;  investigation, K.O., I.K. and E.O.; writing---original draft preparation, K.O.; supervision, K.O.; project administration, K.O.; funding acquisition, K.O. All authors have read and agreed to the published version of the manuscript.}

\funding{The study was supported by the Russian Science Foundation, grant \mbox{No. 22-21-00265},
\url{https://rscf.ru/project/22-21-00265/}}

\conflictsofinterest{The authors declare no conflict of interest. }

\institutionalreview{Not applicable.}

\informedconsent{Not applicable.}

\dataavailability{Not applicable.} 

%

\reftitle{References}


%
\externalbibliography{yes}
\bibliography{OsetrinListOfCitedPublications-2023-06-17,Kirnos-2023-06-17}

\end{document}